\documentclass[a4paper]{article}
\usepackage{fancyvrb}
\usepackage{color}
\makeatletter
\def\PY@reset{\let\PY@it=\relax \let\PY@bf=\relax%
    \let\PY@ul=\relax \let\PY@tc=\relax%
    \let\PY@bc=\relax \let\PY@ff=\relax}
\def\PY@tok#1{\csname PY@tok@#1\endcsname}
\def\PY@toks#1+{\ifx\relax#1\empty\else%
    \PY@tok{#1}\expandafter\PY@toks\fi}
\def\PY@do#1{\PY@bc{\PY@tc{\PY@ul{%
    \PY@it{\PY@bf{\PY@ff{#1}}}}}}}
\def\PY#1#2{\PY@reset\PY@toks#1+\relax+\PY@do{#2}}
\@namedef{PY@tok@w}{\def\PY@tc##1{\textcolor[rgb]{0.73,0.73,0.73}{##1}}}
\@namedef{PY@tok@c}{\let\PY@it=\textit\def\PY@tc##1{\textcolor[rgb]{0.24,0.48,0.48}{##1}}}
\@namedef{PY@tok@cp}{\def\PY@tc##1{\textcolor[rgb]{0.61,0.40,0.00}{##1}}}
\@namedef{PY@tok@k}{\let\PY@bf=\textbf\def\PY@tc##1{\textcolor[rgb]{0.00,0.50,0.00}{##1}}}
\@namedef{PY@tok@kp}{\def\PY@tc##1{\textcolor[rgb]{0.00,0.50,0.00}{##1}}}
\@namedef{PY@tok@kt}{\def\PY@tc##1{\textcolor[rgb]{0.69,0.00,0.25}{##1}}}
\@namedef{PY@tok@o}{\def\PY@tc##1{\textcolor[rgb]{0.40,0.40,0.40}{##1}}}
\@namedef{PY@tok@ow}{\let\PY@bf=\textbf\def\PY@tc##1{\textcolor[rgb]{0.67,0.13,1.00}{##1}}}
\@namedef{PY@tok@nb}{\def\PY@tc##1{\textcolor[rgb]{0.00,0.50,0.00}{##1}}}
\@namedef{PY@tok@nf}{\def\PY@tc##1{\textcolor[rgb]{0.00,0.00,1.00}{##1}}}
\@namedef{PY@tok@nc}{\let\PY@bf=\textbf\def\PY@tc##1{\textcolor[rgb]{0.00,0.00,1.00}{##1}}}
\@namedef{PY@tok@nn}{\let\PY@bf=\textbf\def\PY@tc##1{\textcolor[rgb]{0.00,0.00,1.00}{##1}}}
\@namedef{PY@tok@ne}{\let\PY@bf=\textbf\def\PY@tc##1{\textcolor[rgb]{0.80,0.25,0.22}{##1}}}
\@namedef{PY@tok@nv}{\def\PY@tc##1{\textcolor[rgb]{0.10,0.09,0.49}{##1}}}
\@namedef{PY@tok@no}{\def\PY@tc##1{\textcolor[rgb]{0.53,0.00,0.00}{##1}}}
\@namedef{PY@tok@nl}{\def\PY@tc##1{\textcolor[rgb]{0.46,0.46,0.00}{##1}}}
\@namedef{PY@tok@ni}{\let\PY@bf=\textbf\def\PY@tc##1{\textcolor[rgb]{0.44,0.44,0.44}{##1}}}
\@namedef{PY@tok@na}{\def\PY@tc##1{\textcolor[rgb]{0.41,0.47,0.13}{##1}}}
\@namedef{PY@tok@nt}{\let\PY@bf=\textbf\def\PY@tc##1{\textcolor[rgb]{0.00,0.50,0.00}{##1}}}
\@namedef{PY@tok@nd}{\def\PY@tc##1{\textcolor[rgb]{0.67,0.13,1.00}{##1}}}
\@namedef{PY@tok@s}{\def\PY@tc##1{\textcolor[rgb]{0.73,0.13,0.13}{##1}}}
\@namedef{PY@tok@sd}{\let\PY@it=\textit\def\PY@tc##1{\textcolor[rgb]{0.73,0.13,0.13}{##1}}}
\@namedef{PY@tok@si}{\let\PY@bf=\textbf\def\PY@tc##1{\textcolor[rgb]{0.64,0.35,0.47}{##1}}}
\@namedef{PY@tok@se}{\let\PY@bf=\textbf\def\PY@tc##1{\textcolor[rgb]{0.67,0.36,0.12}{##1}}}
\@namedef{PY@tok@sr}{\def\PY@tc##1{\textcolor[rgb]{0.64,0.35,0.47}{##1}}}
\@namedef{PY@tok@ss}{\def\PY@tc##1{\textcolor[rgb]{0.10,0.09,0.49}{##1}}}
\@namedef{PY@tok@sx}{\def\PY@tc##1{\textcolor[rgb]{0.00,0.50,0.00}{##1}}}
\@namedef{PY@tok@m}{\def\PY@tc##1{\textcolor[rgb]{0.40,0.40,0.40}{##1}}}
\@namedef{PY@tok@gh}{\let\PY@bf=\textbf\def\PY@tc##1{\textcolor[rgb]{0.00,0.00,0.50}{##1}}}
\@namedef{PY@tok@gu}{\let\PY@bf=\textbf\def\PY@tc##1{\textcolor[rgb]{0.50,0.00,0.50}{##1}}}
\@namedef{PY@tok@gd}{\def\PY@tc##1{\textcolor[rgb]{0.63,0.00,0.00}{##1}}}
\@namedef{PY@tok@gi}{\def\PY@tc##1{\textcolor[rgb]{0.00,0.52,0.00}{##1}}}
\@namedef{PY@tok@gr}{\def\PY@tc##1{\textcolor[rgb]{0.89,0.00,0.00}{##1}}}
\@namedef{PY@tok@ge}{\let\PY@it=\textit}
\@namedef{PY@tok@gs}{\let\PY@bf=\textbf}
\@namedef{PY@tok@gp}{\let\PY@bf=\textbf\def\PY@tc##1{\textcolor[rgb]{0.00,0.00,0.50}{##1}}}
\@namedef{PY@tok@go}{\def\PY@tc##1{\textcolor[rgb]{0.44,0.44,0.44}{##1}}}
\@namedef{PY@tok@gt}{\def\PY@tc##1{\textcolor[rgb]{0.00,0.27,0.87}{##1}}}
\@namedef{PY@tok@err}{\def\PY@bc##1{{\setlength{\fboxsep}{\string -\fboxrule}\fcolorbox[rgb]{1.00,0.00,0.00}{1,1,1}{\strut ##1}}}}
\@namedef{PY@tok@kc}{\let\PY@bf=\textbf\def\PY@tc##1{\textcolor[rgb]{0.00,0.50,0.00}{##1}}}
\@namedef{PY@tok@kd}{\let\PY@bf=\textbf\def\PY@tc##1{\textcolor[rgb]{0.00,0.50,0.00}{##1}}}
\@namedef{PY@tok@kn}{\let\PY@bf=\textbf\def\PY@tc##1{\textcolor[rgb]{0.00,0.50,0.00}{##1}}}
\@namedef{PY@tok@kr}{\let\PY@bf=\textbf\def\PY@tc##1{\textcolor[rgb]{0.00,0.50,0.00}{##1}}}
\@namedef{PY@tok@bp}{\def\PY@tc##1{\textcolor[rgb]{0.00,0.50,0.00}{##1}}}
\@namedef{PY@tok@fm}{\def\PY@tc##1{\textcolor[rgb]{0.00,0.00,1.00}{##1}}}
\@namedef{PY@tok@vc}{\def\PY@tc##1{\textcolor[rgb]{0.10,0.09,0.49}{##1}}}
\@namedef{PY@tok@vg}{\def\PY@tc##1{\textcolor[rgb]{0.10,0.09,0.49}{##1}}}
\@namedef{PY@tok@vi}{\def\PY@tc##1{\textcolor[rgb]{0.10,0.09,0.49}{##1}}}
\@namedef{PY@tok@vm}{\def\PY@tc##1{\textcolor[rgb]{0.10,0.09,0.49}{##1}}}
\@namedef{PY@tok@sa}{\def\PY@tc##1{\textcolor[rgb]{0.73,0.13,0.13}{##1}}}
\@namedef{PY@tok@sb}{\def\PY@tc##1{\textcolor[rgb]{0.73,0.13,0.13}{##1}}}
\@namedef{PY@tok@sc}{\def\PY@tc##1{\textcolor[rgb]{0.73,0.13,0.13}{##1}}}
\@namedef{PY@tok@dl}{\def\PY@tc##1{\textcolor[rgb]{0.73,0.13,0.13}{##1}}}
\@namedef{PY@tok@s2}{\def\PY@tc##1{\textcolor[rgb]{0.73,0.13,0.13}{##1}}}
\@namedef{PY@tok@sh}{\def\PY@tc##1{\textcolor[rgb]{0.73,0.13,0.13}{##1}}}
\@namedef{PY@tok@s1}{\def\PY@tc##1{\textcolor[rgb]{0.73,0.13,0.13}{##1}}}
\@namedef{PY@tok@mb}{\def\PY@tc##1{\textcolor[rgb]{0.40,0.40,0.40}{##1}}}
\@namedef{PY@tok@mf}{\def\PY@tc##1{\textcolor[rgb]{0.40,0.40,0.40}{##1}}}
\@namedef{PY@tok@mh}{\def\PY@tc##1{\textcolor[rgb]{0.40,0.40,0.40}{##1}}}
\@namedef{PY@tok@mi}{\def\PY@tc##1{\textcolor[rgb]{0.40,0.40,0.40}{##1}}}
\@namedef{PY@tok@il}{\def\PY@tc##1{\textcolor[rgb]{0.40,0.40,0.40}{##1}}}
\@namedef{PY@tok@mo}{\def\PY@tc##1{\textcolor[rgb]{0.40,0.40,0.40}{##1}}}
\@namedef{PY@tok@ch}{\let\PY@it=\textit\def\PY@tc##1{\textcolor[rgb]{0.24,0.48,0.48}{##1}}}
\@namedef{PY@tok@cm}{\let\PY@it=\textit\def\PY@tc##1{\textcolor[rgb]{0.24,0.48,0.48}{##1}}}
\@namedef{PY@tok@cpf}{\let\PY@it=\textit\def\PY@tc##1{\textcolor[rgb]{0.24,0.48,0.48}{##1}}}
\@namedef{PY@tok@c1}{\let\PY@it=\textit\def\PY@tc##1{\textcolor[rgb]{0.24,0.48,0.48}{##1}}}
\@namedef{PY@tok@cs}{\let\PY@it=\textit\def\PY@tc##1{\textcolor[rgb]{0.24,0.48,0.48}{##1}}}

\def\PYZus{\char`\_}

\def\PYZpc{\char`\%}

\makeatother

\usepackage{hyperref}
\usepackage{amssymb}
\usepackage{graphicx}
\usepackage[capitalise,noabbrev]{cleveref}
\usepackage{subcaption}
\usepackage{orcidlink}
\newcommand{\R}{\mathbb{R}}

\newcommand{\data}{\ensuremath\mathcal{A}}

\title{Fast persistent homology computation\\ for functions on~$\mathbb{R}$}
\author{Marc Glisse\,\orcidlink{0000-0001-6914-1651}\footnote{\href{mailto:marc.glisse@inria.fr}{\texttt{marc.glisse@inria.fr}}. Université Paris-Saclay, CNRS, Inria, Laboratoire de Mathématiques d'Orsay, 91405, Orsay, France.}}
\begin{document}
\maketitle
\begin{abstract}
  0-dimensional persistent homology is known, from a computational point of view, as the easy case. Indeed, given a list of $n$ edges in non-decreasing order of filtration value, one only needs a union-find data structure to keep track of the connected components and we get the persistence diagram in time $O(n\alpha(n))$. The running time is thus usually dominated by sorting the edges in $\Theta(n\log(n))$. A little-known fact is that, in the particularly simple case of studying the sublevel sets of a piecewise-linear function on $\mathbb{R}$ or $\mathbb{S}^1$, persistence can actually be computed in linear time. This note presents a simple algorithm that achieves this complexity and an extension to image persistence. An implementation is available in Gudhi~\cite{gudhi:urm}.
\end{abstract}

\section{Main idea}
The piecewise-linear (PL) function $f:\R\rightarrow\R$ is defined by its image at each vertex, represented as an array $\data$. Usual algorithms for sublevel set persistent homology first replace this with an equivalent lower-star filtration defined on a path graph (which is a special case of simplicial complex and cubical complex). However, this is not needed for our approach which works at the level of PL functions until the end.

We call a \emph{pattern} several consecutive elements in an array whose values have the same order as the name of the pattern. For instance, if $\data[i+1]<\data[i+2]<\data[i]$, we have a pattern 312.

Note that if we have two identical consecutive values (pattern 11), it is fine to keep only one, this does not affect the persistence diagram. Also, for two (not necessarily consecutive) identical values, the stability theorem tells us that it is fine to simulate simplicity by assuming for instance that the second one is larger, so we do not need to consider patterns like 121 but only 132 or 231. The last trivial remark is that in a pattern 123, we can drop the 2. In particular, we only need to handle sequences of alternating local minima and local maxima.

\begin{figure}
  \centering
  \includegraphics{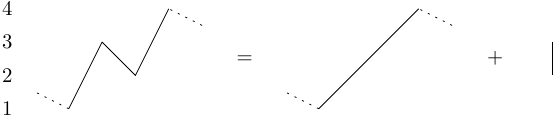}
  \caption{Pattern 1324 and its reduction.}
  \label{fig:1324}
\end{figure}
The main ingredient is that when we see a pattern 1324 (or its symmetric 4231), the elements represented by 2 and 3 in the pattern define a pair in the persistence diagram. Indeed, looking at the sublevel sets, a connected component appears at 2, and at 3 it merges with an older component that has existed since at least 1. We can thus remove those two elements, 4 now directly follows 1
, and the persistence diagram of the reduced array plus the pair $(2,3)$ is equal to the persistence diagram of the original array, see \cref{fig:1324}..

\begin{figure}
  \centering
  \includegraphics{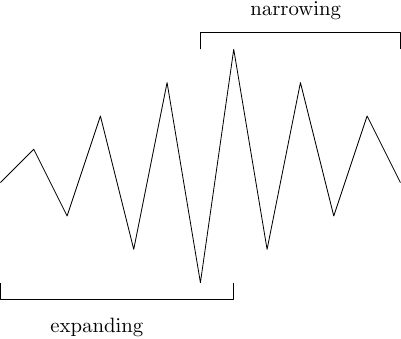}
  \caption{A 2-phase sequence.}
  \label{fig:2phase}
\end{figure}
After reducing the sequence based on those remarks, we are left with a sequence with no 123, 321, 1324 or 4231 pattern. It is easy to see that such a sequence has a very specific \emph{2-phase} shape depicted in \cref{fig:2phase}: it alternates between local minima and local maxima, in a first \emph{expanding} phase the maxima are increasing and the minima decreasing, and in a second \emph{narrowing} phase the maxima are decreasing and the minima increasing (of course one of the phases may be empty). Indeed, as soon as we see a pattern 132, the next element can be neither lower than 2 (pattern 321) nor higher than 3 (pattern 1324), so the triple that follows and shares 2 elements with 132 can only follow the pattern 312. Similarly, 312 can only be followed by 132, and the narrowing pattern only stops at the end of the sequence. The expanding phase is symmetric to the narrowing one, and we only need to notice that 2 consecutive local minima must be in a pattern 132 or 231 to conclude.

The extremities also provide some opportunities (computing extended persistence~\cite{extended-pers} would require small tweaks here). For instance if the sequence starts with 21, we can remove 2. If it starts with 231, we can pair off 2 and 3, remove them and start the sequence at 1. Symmetric operations are possible at the other extremity. This is sufficient to reduce a 2-phased sequence as obtained above to just one point, the global minimum, and we have the whole persistence diagram.

To apply these operations efficiently, we propose adding the values one by one from left to right, maintaining a reduced sequence with the invariant that it has a narrowing pattern and starts by 12 (unless it is reduced to a single value). This invariant is equivalent to the absence of patterns 123, 321, 1324 and 4231, and of patterns 21 and 231 at the beginning. With every new value, we check if appending it breaks the invariant, or equivalently if a simplification involving the new element is possible, and we simplify until the invariant is restored. After processing the whole input, and after removing the last element in case the reduced sequence ends in 12, we can just pair and remove the last 2 elements (terminating 132 pattern) recursively until only 1 element remains. Since we only go back when removing elements, the amortized complexity per element is constant and the whole algorithm takes linear time.

\section{Function on a circle}
For a PL function defined on the circle $\mathbb{S}^1$, we do not have extremities at which we could simplify, but that is unnecessary, since removing the patterns 123/321 and 1324/4231 is sufficient to get down to just 2 values. Indeed, without 123 and 321, the sequence has an even length and alternates between local minima and maxima. If the sequence has length at least 4, 2 consecutive minima are involved in a pattern 132 or 231. By symmetry, we assume there is a pattern 132. As in the line case, when a narrowing pattern starts, it cannot end until the end of the sequence. However, on a circle, the sequence is periodic and does not end. In particular, it reaches this 1 again. In a narrowing sequence, the minima increase, so when we reach 1 again, it must be larger than 2, a contradiction.

From an algorithmic point of view, we can cut the circle at an arbitrary point, make a first pass to get a 2-phase pattern, then reconnect the 2 extremities and simplify at the junction until we have only 2 values left. It may be convenient to use the global minimum (or maximum, or both) as initial cutting point, although it does not change the linear complexity.

\section{Parallelism}
\begin{figure}
  \centering
  \includegraphics[page=1]{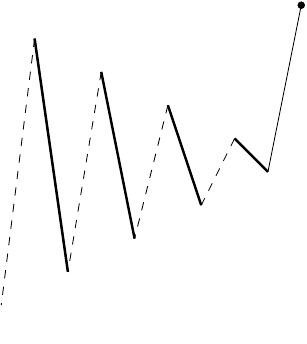}
  \hfill
  \includegraphics[page=2]{nonlocal}
  \caption{Non-locality: the last element may determine the pairing of all the other points.}
  \label{fig:nonlocal}
\end{figure}

Although we expect this approach is fast enough in practice not to require parallelization, it is tempting to try it. We can split the segment into smaller segments, simplify each of them to a 2-phase shape in parallel, and collect the pairs the simplifications find. And we can iterate, possibly using the point where the phase changes (the minimum) inside each segment as the new splitting points, or merging adjacent segments. For a fairly nice function, this should reduce the sequence significantly and let us finish sequentially. However, if for instance the input already has a narrowing shape from the beginning, the parallel phases will do almost nothing. This is then just a heuristic. There is little hope of a perfectly parallel algorithm because of the non-locality of persistent homology, \cref{fig:nonlocal} shows that for a narrowing shape, adding at the end a very large or very small value can change the pairing of all the points.

An extreme version of this parallelization (with overlapping segments of length 4) on GPU is described in \cite{glisse:LIPIcs.ESA.2026.5}.

\section{Experiments}
\begin{figure}
\begin{Verbatim}[commandchars=\\\{\}]
\PY{n}{In} \PY{p}{[}\PY{l+m+mi}{1}\PY{p}{]}\PY{p}{:} \PY{k+kn}{import} \PY{n+nn}{numpy} \PY{k}{as} \PY{n+nn}{np}
   \PY{o}{.}\PY{o}{.}\PY{o}{.}\PY{p}{:} \PY{k+kn}{from} \PY{n+nn}{gudhi}\PY{n+nn}{.}\PY{n+nn}{sklearn}\PY{n+nn}{.}\PY{n+nn}{cubical\PYZus{}persistence} \PY{k+kn}{import} \PY{o}{*}
   \PY{o}{.}\PY{o}{.}\PY{o}{.}\PY{p}{:} \PY{n}{cp} \PY{o}{=} \PY{n}{CubicalPersistence}\PY{p}{(}\PY{n}{homology\PYZus{}dimensions}\PY{o}{=}\PY{l+m+mi}{0}\PY{p}{)}

\PY{n}{In} \PY{p}{[}\PY{l+m+mi}{2}\PY{p}{]}\PY{p}{:} \PY{o}{\PYZpc{}}\PY{k}{time} fun = np.random.rand(1\PYZus{}000\PYZus{}000\PYZus{}000)
\PY{n}{CPU} \PY{n}{times}\PY{p}{:} \PY{n}{user} \PY{l+m+mf}{4.63} \PY{n}{s}\PY{p}{,} \PY{n}{sys}\PY{p}{:} \PY{l+m+mi}{479} \PY{n}{ms}\PY{p}{,} \PY{n}{total}\PY{p}{:} \PY{l+m+mf}{5.1} \PY{n}{s}
\PY{n}{Wall} \PY{n}{time}\PY{p}{:} \PY{l+m+mf}{5.1} \PY{n}{s}

\PY{n}{In} \PY{p}{[}\PY{l+m+mi}{3}\PY{p}{]}\PY{p}{:} \PY{o}{\PYZpc{}}\PY{k}{time} diags = cp.fit\PYZus{}transform([fun])
\PY{n}{CPU} \PY{n}{times}\PY{p}{:} \PY{n}{user} \PY{l+m+mf}{9.1} \PY{n}{s}\PY{p}{,} \PY{n}{sys}\PY{p}{:} \PY{l+m+mi}{979} \PY{n}{ms}\PY{p}{,} \PY{n}{total}\PY{p}{:} \PY{l+m+mf}{10.1} \PY{n}{s}
\PY{n}{Wall} \PY{n}{time}\PY{p}{:} \PY{l+m+mf}{10.1} \PY{n}{s}
\end{Verbatim}
\caption{Example code.}
  \label{fig:code}
\end{figure}

%
%

An implementation for the line case is available in Gudhi~\cite{gudhi:urm}. As an experiment, see \cref{fig:code}, we generated a random NumPy array of size $10^9$, which took 5 seconds, then computed its persistence in 10 seconds (NumPy would take 90 seconds to sort the array). The random case is not very favorable, a monotonic function takes only 3 seconds, and a constant function 1.4 seconds. As an other point of comparison, copying the whole array already takes 1 second. Computing persistence of a 1D function can be considered fast enough at this point.

\section{Extension: image persistence}

Assume we now have 2 PL functions $f,g:\R\rightarrow\R$ such that $\forall x\in\R: f(x)\leq g(x)$. We consider the sequence $\left(\textrm{Im } H\left(g^{-1}((-\infty,t])\right)\rightarrow H\left(f^{-1}((-\infty,t])\right)\right)_{t\in\R}$. Inclusions between sublevel sets naturally induce morphisms between the spaces of this sequence, which can be seen as a persistence module called image persistence~\cite{pers-im-ker}.

\begin{figure}
  \begin{subfigure}{.45\textwidth}
  \includegraphics[width=\textwidth]{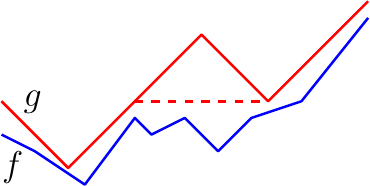}
  \end{subfigure}
  \hfill
  \begin{subfigure}{.45\textwidth}
  \includegraphics[width=\textwidth]{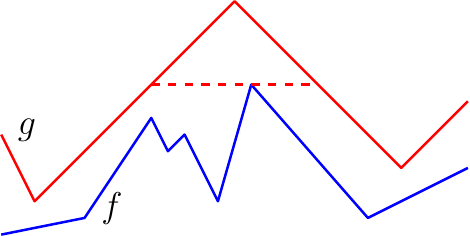}
  \end{subfigure}
  \caption{2 consecutive local minima of $g$.}
  \label{fig:im2min}
\end{figure}
\begin{figure}
  \begin{subfigure}{.45\textwidth}
  \includegraphics[width=\textwidth]{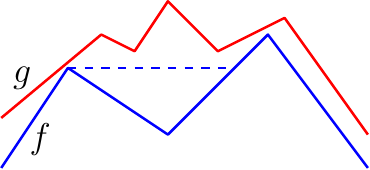}
  \end{subfigure}
  \hfill
  \begin{subfigure}{.45\textwidth}
  \includegraphics[width=\textwidth]{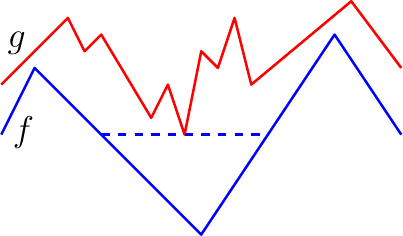}
  \end{subfigure}
  \caption{2 consecutive local maxima of $f$.}
  \label{fig:im2max}
\end{figure}

While with a single function we had local minima creating connected components and local maxima merging them, the roles are now split between the two functions: the connected components are created by the local minima of $g$, while their merging is determined by the local maxima of $f$ (merging can thus happen \emph{before} a component is even born, in which case the component never exists in the image).
To be a bit more formal, we define some transformations that preserve image persistence and eventually reach the case $f=g$ that was solved in previous sections.

\paragraph{Between minima.}
Refer to \cref{fig:im2min}.
Consider 2 consecutive local minima of $g$: $g(t_1)=m_1$ and $g(t_2)=m_2$. Without loss of generality, we can assume that $t_1<t_2$ and $m_1<m_2$. Let $m_f$ be the maximum value reached by $f$ on $[t_1,t_2]$. We can decrease $g$ so that it does not exceed $m=\max(m_2, m_f)$ on $[t_1,t_2]$ without changing the image filtration, since $t_1$ and $t_2$ are already connected in $f^{-1}((-\infty,m])$. If $m_f\leq m_2$, this removes one local minimum of $g$.

\paragraph{Between maxima.}
Refer to \cref{fig:im2max}.
Consider 2 consecutive local maxima of $f$: $f(t_1)=m_1$ and $f(t_2)=m_2$. Without loss of generality, we can assume that $t_1<t_2$ and $m_1<m_2$. Let $m_g$ be the minimum value reached by $g$ on $[t_1,t_2]$. We can increase $f$ so that it does not go below $m=\min(m_1, m_g)$ on $[t_1,t_2]$ without changing the image filtration, since for $u<m$ any connected component of $f^{-1}((-\infty,u])\cap [t_1,t_2]$ is disconnected from $(-\infty,t_1)\cup(t_2,\infty)$ and $g^{-1}((-\infty,u])\cap [t_1,t_2]=\emptyset$ has nothing to send into it (that component is in the cokernel). If $m_g\geq m_1$, this removes one local maximum of $f$.

~

Using those 2 transformations as much as possible, we are left with a sequence of local minima of $g$ and local maxima of $f$, where we cannot have 2 consecutive elements of the same type (for instance if we have 2 minima of $g$ without a maximum of $f$ in between, we must be in the case $m_f\leq m_2$ and we can remove a minimum), where each maximum is larger than each adjacent minimum (again the case $m_f\leq m_2$ or $m_g\geq m_1$), and each local maximum of $f$ is also the maximum of $g$ between the adjacent local minima of $g$ (and symmetrically for minima of $g$), so $f$ and $g$ match (I skipped explaining how to handle the extremities, but there is no particular complexity there). This reduction to the case of a single function takes linear time.

Let me mention a tempting but wrong alternate approach to the problem. We could create a vertex for each local minimum of $g$ with filtration value this minimum, connect 2 adjacent vertices with an edge with filtration value the maximum of $f$ on this interval (or just the maximum of the adjacent vertices if $f$ is too low), and compute the persistence diagram of this filtered graph. However, this ignores the fact that while a maximum of $f$ on $[t_2,t_3]$ may be too low to matter on this interval, it may be very relevant when looking at a larger interval $[t_1,t_4]$.

Future work could include computing (co-)kernel persistence, or even looking at this in a multi-parameter setting (the second parameter selects between $f$ and $g$). Giving an interpretation in terms of \emph{windows} \cite{Biswas2023} could also be interesting.

\section{Lower bound for trees}
\begin{figure}
  \centering
  \includegraphics{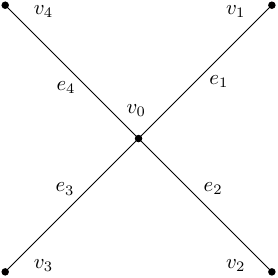}
  \caption{Tree construction for the lower bound.}
  \label{fig:borneinf}
\end{figure}
After lines (or circles), a natural next case to consider is trees. Unlike general graphs, for trees we already know that every edge connects two different connected components, so union-find can be implemented in linear time~\cite{unionfindtreelinear}. However, even in this favorable case, we cannot escape the need for sorting. Considering $n$ numbers $x_1,\ldots,x_n$ in $(0,1)$, we assign filtration values to the tree of \cref{fig:borneinf} as $f(e_i)=-f(v_i)=x_i$ and $f(v_0)=0$. This is a filtration, since the edges have positive value and the vertices non-positive. If the values $x_i$ are sorted: $x_1<x_2<\ldots<x_n$, we have $f(v_n)<\ldots<f(v_1)<f(v_0)<f(e_1)<\ldots<f(e_n)$. The persistence diagram consists of the pairs $(v_i,e_{i+1})$, leaving $v_n$ unpaired as the global minimum (one way to see it is that $(v_0,e_1)$ forms a shallow pair~\cite{edelsbrunner2026posetcancellationsinducedgradient}, and cancelling it yields a similar tree with $v_1$ in the center). While the unsorted list of persistence pairs does not directly provide the sorted list of $x_i$ (unless one cheats with a hash table), it is as hard to compute, there are $n!$ possible pairings (in bijection with the permutations), which require $\Omega(n\log n)$ comparisons so we can distinguish them in the comparison model.

\section{Related work}
Several papers have appeared around the same time as this one. In~\cite{Biswas2023}, the authors characterize the persistent homology of time series with a notion of \emph{window}, related to the patterns that could appear after some simplifications. In~\cite{banana24,banana25}, using the same notion of window, the authors design a structure called \emph{banana tree} that can be maintained efficiently (except in pathological cases like \cref{fig:nonlocal}) under dynamic modifications of the input time series, and from which they can extract the extended persistence diagram and some information related to the merge tree. As a special case, it also allows them to compute a static persistence diagram in linear time, although their algorithm is significantly more complicated than the one presented here.

When we only have alternating local minima and maxima, we can interpret the minima as vertices and the maxima as edges. The patterns considered here then correspond to \emph{apparent pairs}~\cite{bauer_et_al:LIPIcs.SoCG.2024.15} or \emph{shallow pairs}~\cite{edelsbrunner2026posetcancellationsinducedgradient,edelsbrunner_et_al:LIPIcs.SoCG.2026.41}. Following the presentation in \cite{edelsbrunner2026posetcancellationsinducedgradient}, removing a pattern corresponds to cancelling the shallow pair, and in this special case the resulting Lefschetz complex remains a 1-dimensional cubical complex. Besides, they notice that the only case where there is no shallow pair is when the boundary matrix is zero, i.e. in our case there is a single vertex left. This provides an alternate proof of the validity of the algorithm presented here.

\bibliographystyle{plainurl}
\bibliography{biblio}

\end{document}